# Synchronization in a market model with time delays

Ghassan Dibeh[1]
Department of Economics
Lebanese American University
Chouran Street
Beirut 1102 2801
Lebanon
gdibeh@lau.edu.lb

Omar El Deeb
Mathematics Institute
University of Warwick
Coventry, United Kingdom

## Abstract

We examine a system of $N = 2$ coupled non-linear delay-differential equations representing financial market dynamics. In such time delay systems, coupled oscillations have been derived. We linearize the system for small time delays and study its collective dynamics. Using analytical and numerical solutions, we obtain the bifurcation diagrams and analyze the corresponding regions of amplitude death, phase locking, limit cycles and market synchronization in terms of the system frequency-like parameters and time delays. We further numerically explore higher order systems with $N > 2$, and demonstrate that limit cycles can be maintained for coupled $N-$asset models with appropriate parameterization.

Keywords: Speculative markets; Synchronization; time delays; limit cycles.

## 1. Introduction

---

[1] Corresponding author.

System synchronization dynamics in the realm of physics manifest in a multitude of physical systems, where diverse components or elements tend to align their behaviors over time. Notable instances include coupled oscillators, such as pendulum clocks on a common wall, synchronizing their swings through phase synchronization whereas celestial mechanics reveal planetary orbit synchronization, like the Moon's tidal locking to the Earth. In lasers, photons synchronize their phases to produce coherent light emissions. Superconductors exhibit synchronization as electrons pair and move without resistance and magnetic materials can synchronize atomic spins, leading to phenomena like ferromagnetism. In biophysics, the synchronization of heart muscle contractions and pumping frequencies is vital. Synchronization is also essential in neural networks, where it refers to the coordinated activity of neurons or units within the network, where their firing patterns align over time [1,2]. It allows different parts of the network to work together harmoniously, facilitating the emergence of coherent representations and patterns in the learned data.

In financial market dynamics, spillovers and correlations between different markets or assets are prevalent in modern financial markets [3-5]. In addition, in cases of extreme events such as market crashes, contagion between different markets has been observed throughout different financial crises such as the Asian financial crisis of 1997 and the Great Financial Crisis (GFC) of 2008. In this respect, models of speculative dynamics that exhibit financial cycles and oscillations in single markets have been developed [6-7] with extensions to include coupling between markets leading to synchronization between financial or asset markets [8]. The delay-differential equations model presented in [8] shows that two-markets that are coupled through prices exhibit limit cycles. These limit cycles can be seen as the result of a spillover or contagion effect between the two markets. Further analysis of this model has shown that Hopf bifurcations exist analytically [9,10].

This paper extends the 2-asset market model generalizing the model to include the existence of delays in the contagion process. Using analytical and numerical solutions, we obtain the bifurcation diagrams and analyze the corresponding regions of amplitude death, phase locking and limit cycles in terms of the system frequency-like parameters and time delays. Moreover, the model is extended to higher dimensions (N=4) which would represent more closely real-world financial markets where more than two assets are present. In addition, the model is extended to a generic N-assets market. These extensions allow the study of rise of synchronization in N-asset markets in time delayed models in the same vein as models studied in physical systems [11, 12].

The paper is divided as follows: Section two introduces the model. Section three analyzes the model for synchronization with short time delays. Section four presents a general treatment for the market system deriving the conditions for the existence of amplitude death and limit cycle regions. Section five presents a numerical simulation of an N=4 system verifying the onset of limit cycles. Section six introduces a generic N-dimensional model that allows analytical investigation of amplitude death regions for systems with identical frequencies. Section seven concludes.

## 2. The model

We consider the interaction of a large number $N$ of asset markets that are globally coupled with linear time delayed couplings. We describe such a system with the following set of coupled delay differential equations:

$$\dot{p}_i = (1 - m_i) \tanh[p_i(t) - p_i(t - \tau_i)] p_i(t) - m_i(p_i(t) - v_i)p_i(t)$$
$$+ K_{ij} \sum_{j=1, j \neq i}^{N} \left(p_j(t - \tau_j) - p_i(t)\right) \quad (1)$$

The last term is the coupling term of market $i$ with all other markets $j \neq i$, and $K_{ij}$ is the coupling constant between markets $i$ and $j$.

This is a highly non-linear model due to the tanh dependence, and analytical solution are not available.

The tanh function is an activation function commonly used in artificial neural networks. It has a range between -1 and 1, and it is centered around zero. In neural networks, the tanh activation function is often used in hidden layers to introduce non-linearity into the model, enabling the network to learn complex patterns and representations [1,2]. In our model, system (1) can be linearized and its dynamics could be studied analytically and numerically.

To analytically prove the existence of phase locking regions, we analyze a two-asset market ($N = 2$), without loss of generality. System (1) reduces to:

$$\begin{cases} \dot{p}_1 = (1 - m_1) \tanh[p_1(t) - p_1(t - \tau_1)] p_1(t) - m_1(p_1(t) - v_1)p_1(t) + K_{12}\left(p_2(t - \tau_2) - p_1(t)\right) \\ \dot{p}_2 = (1 - m_2) \tanh[p_2(t) - p_2(t - \tau_2)] p_2(t) - m_2(p_2(t) - v_2)p_2(t) + K_{21}\left(p_1(t - \tau_1) - p_2(t)\right) \end{cases} \quad (2)$$

## 3. Synchronization for short time delays

In this part, we analytically and numerically analyze the system in presence of short time delays between two markets, and we show how a pattern of limit cycles emerges for the price dynamics. Mathematically, for short time delays, eq. (2) could be simplified using a Taylor series expansion such that $p_i(t - \tau_i) = p(t) - \tau_i \dot{p}_i(t)$. In this case, the fixed points of the system distinguished by $\dot{p}_i = 0$ would be determined by the algebraic system:

$$\begin{cases} -m_1(p_1(t) - v_1)p_1(t) + K_{12}\left(p_2(t) - p_1(t)\right) = 0 \\ -m_2(p_2(t) - v_2)p_2(t) + K_{21}\left(p_1(t) - p_2(t)\right) = 0 \end{cases} \quad (3)$$

Eq. (3) can be solved analytically in terms of the system parameters, and it reveals the presence of four pairs of solution: $(p_1, p_2) = (0,0)$, $(p_1, p_2) = (p_1^*, p_2^*)$ where $p_1^*$ and $p_2^*$ are positive valued real numbers expressed in terms of $m_1, m_2, v_1, v_2$ and $K$ (while assuming $K_{12} = K_{21} = K$), and two other complex solutions for $p_i$. Limit cycles can be immediately excluded for the complex solutions as they don't correspond to actual prices. In addition, limit cycles for prices are forbidden around the fixed point (0,0) since any closed orbit in the phase plane around this point needs necessarily to pass through negative values of $p_i$.

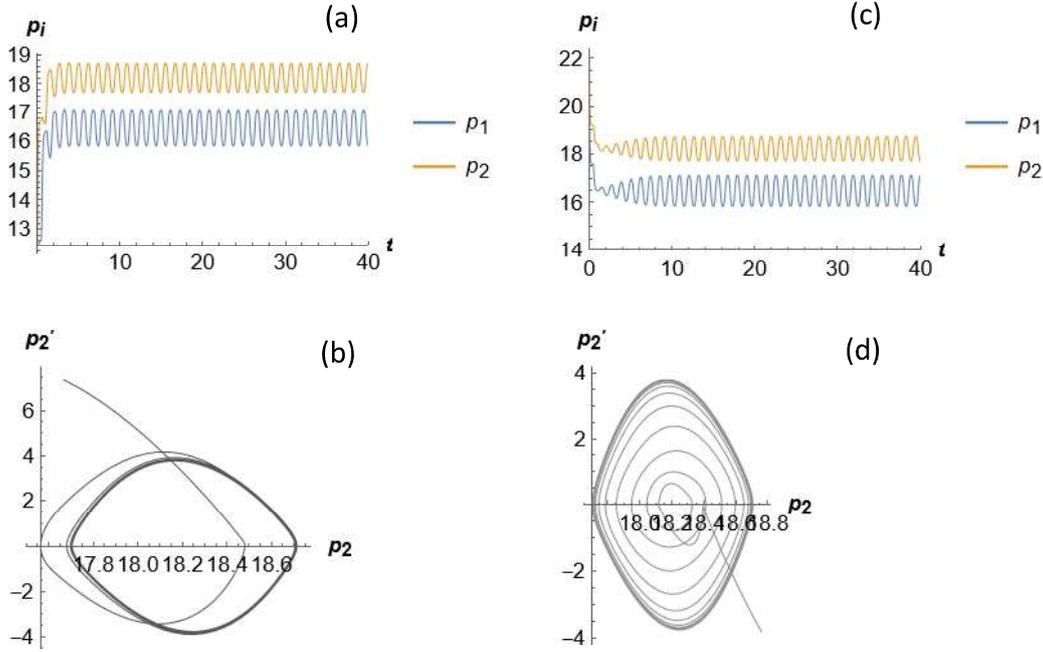

*Figure 1: Plots in the short time delay regime. Plot (a) starts at initial condition well below limit cycle equilibrium with $(p_1(0) = 12, p_2(0) = 10)$, while (c) starts at a higher price $(p_1(0) = 20, p_2(0) = 22)$, but both evolve towards same equilibrium around $(p_1^*, p_2^*) = (16.4, 18.0)$. Plot (b) and (d) represent their $(\dot{p}_2, p_2)$ phase plots respectively.*

Hence, limit cycles could only exist around the solution of the form $(p_1, p_2) = (p_1^*, p_2^*)$, for some specific values that lead to an unstable equilibrium point with closed phase-space orbits around it. In each market, prices in the neighborhood of this fixed point would oscillate around the equilibrium price in real time and will synchronize with the other market dynamics.

This could be visualized in the following example. Consider a two-asset market represented by eq. (2) with parameters: $m_1 = 0.50, m_2 = 0.65, v_1 = 15.0, v_2 = 19.2$ and $K = 7$. We can show that, aside from the 3 excluded equilibria of the system, the solution $(p_1^*, p_2^*) = (16.5, 18.2)$ is a fixed point around which limit cycles and market synchronization take place. Notice that in Fig. (1-a), starting with lower initial conditions/prices $(p_1(0) = 12.0, p_2(0) = 10.0)$, and using the parameters listed above with $\tau_1 = \tau_2 = 0.50$, prices increase towards the equilibrium point $(p_1^*, p_2^*)$, oscillating periodically and synchronizing between the two markets. Similarly in Fig. (1-c), starting with higher initial conditions/prices ($p_1(0) = 20.0$, $p_2(0) = 22.0$), the prices drop, oscillate then continue in limit cycle dynamics around ($p_1^*, p_2^*$),

The phase-plane diagrams show the limit cycle behavior of the system in both cases, falling towards a closed cycle stability orbit in Fig. (1-b), and expanding towards a closed cycle stability orbit in Fig. (1-d), in correspondence to the timeseries plots of $p_2(t)$ above. A similar phase space plot exists for $p_1(t)$ but it was not explicitly displayed in this figure.

Even when starting with initial conditions far away from ($p_1^*, p_2^*$), like prices near the origin, our simulations reveal that the system moves away from the origin as an unstable fixed point until it finally reaches a supercritical limit cycle around ($p_1^*, p_2^*$) for this parameterization. This shows that, once the conditions of existence of limit cycle orbits exist, there would be a flow of solutions towards this orbit, regardless of the initial starting price.

## 4. A general treatment of a two-market system

Our aim at this stage is to study the dynamics of this system in the presence of a generic time delay. Looking at eq. (2), we can realize that $p_i = 0$ is an equilibrium point of the system (among others). We start by analyzing the interesting dynamics around this point, before considering other equilibria. We linearize the system around $p_i = 0$ in order to obtain the characteristic equation:

$$\det(A - \lambda I) = 0 \tag{4}$$

where $A$ is the linearized matrix of system (2) and $I$ is the identity matrix. The perturbations are assumed to have time dependence proportional to $e^{\lambda t}$. Upon linearizing (2), and taking $K_{12} = K_{21} = K$, we get that

$$A = \begin{pmatrix} m_1 v_1 - K & K e^{-\lambda \tau_2} \\ K e^{-\lambda \tau_1} & m_2 v_2 - K \end{pmatrix} \tag{5}$$

Solving eq. (4) we get that:

$$(m_1 v_1 - K - \lambda)(m_2 v_2 - K - \lambda) - K^2 e^{-\lambda(\tau_1 + \tau_2)} = 0 \tag{6}$$

leading to

$$\lambda^2 + 2(K - \overline{\omega})\lambda + b + K^2 e^{-\lambda(\tau_1 + \tau_2)} = 0 \tag{7}$$

where we denote the new parameters by: $\omega_1 = m_1 v_1$, $\omega_2 = m_2 v_2$, $\overline{\omega} = \frac{\omega_1 + \omega_2}{2}$, $\Delta = |\omega_1 - \omega_2|$ and $b = K^2 + \overline{\omega} - \frac{\Delta^2}{4} - 2K\overline{\omega}$.

By substituting the values of $\alpha$ and $\beta$, where they are real numbers, into the equation $\lambda = \alpha + i\beta$, the region where $\alpha < 0$ represents the amplitude death region. The critical curves or marginal stability curves are determined by setting $\alpha = 0$, which means $\lambda$ becomes purely imaginary ($\lambda = i\beta$). By plugging this into (7), the equations that define the critical curves can be derived. We then obtain:

$$-\beta^2 - 2i\beta(K - \overline{\omega}) + b + K^2 e^{-i\beta(\tau_1 + \tau_2)} = 0 \tag{8}$$

which leads to the following conditions:

$$\begin{cases} b - \beta^2 + K^2 \cos(\beta(\tau_1 + \tau_2)) = 0 \\ -2\beta(K - \overline{\omega}) - K^2 \sin(\beta(\tau_1 + \tau_2)) = 0 \end{cases} \tag{9}$$

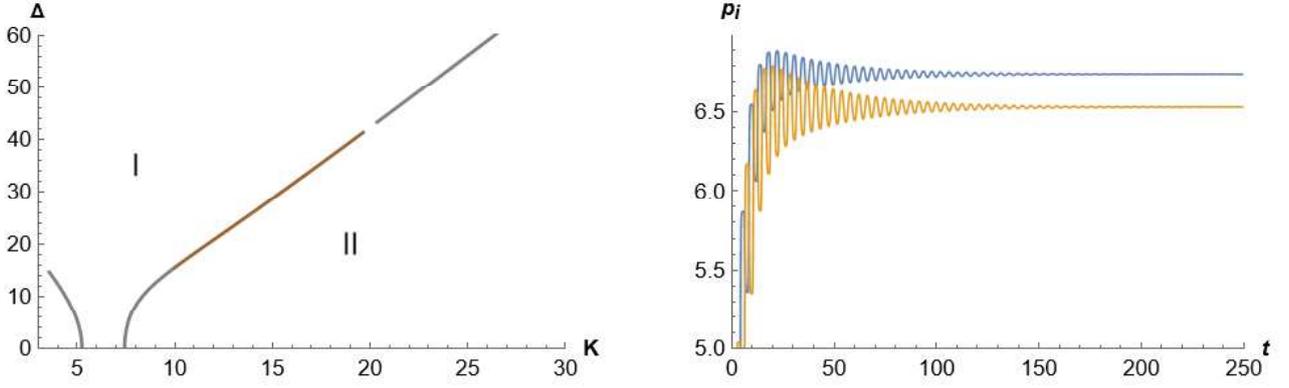

Figure 2: The critical curves in the $(\Delta, K)$ plane on the left show the bifurcation lines that separate a limit cycle region (I) and an amplitude death region (II). The plot on the right represents an amplitude death solution for $\tau_1 = \tau_2 = 2$ and $\overline{\omega} = 5$, corresponding to cases in region (II).

By plotting the $(\Delta, K)$ critical curves, as a parametric plot in terms of the parameter $\beta$ while fixing the other parameters, we could determine the phase locking and amplitude death regions for these specific parameters.

We analyze the limiting case with $\tau_i \to 0$, corresponding to no time delay in the model. The system (9) becomes:

$$\begin{cases} b - \beta^2 + K^2 = 0 \\ -2\beta(K - \overline{\omega}) = 0 \end{cases} \tag{10}$$

whose solutions are clearly given by $K = \overline{\omega}$ or $\beta = 0$ and $\Delta^2 = 4(2K^2 - 2\overline{\omega}K + \overline{\omega}^2)$. The intersection of these curves in the $(\Delta, K)$ plane gives the point of Hopf bifurcation that divides the plane into phase locking and amplitude death regions. Here the intersection occurs at point $(\overline{\omega}, 2\overline{\omega})$. This means that conditions for phase locking or limit cycles are satisfied for $\Delta > 2\overline{\omega}$. However, since $\Delta = |\omega_1 - \omega_2|$, this condition cannot be algebraically satisfied for positive values of both $\omega_1$ and $\omega_2$, hence we prove that there exist no limit cycles for this model in absence of time delays.

In presence of a generic time delay $\tau \neq 0$, we solve system (9) by taking the second equation and substituting, $F = \frac{2\beta}{\sin\beta(\tau_1+\tau_2)}$ we get that:

$$K^2 + FK + \overline{\omega}F = 0 \tag{11}$$

whose roots are given by the equations:

$$K_{\pm} = -F \pm \frac{\sqrt{F^2 + 4\overline{\omega}F}}{2} \tag{12}$$

Finally, substituting (11) in the first equation of (9), we obtain a transcendental equation expressing $\Delta$ in terms of $K$, that allows us to determine regions of amplitude death or frequency locking. This equation is given by:

$$\Delta^2 = 4(K - \overline{\omega})^2 - 4\beta^2 + 4K^2 \cos(\beta(\tau_1 + \tau_2)) \tag{13}$$

This relation between $\Delta$ and $K$ represents the critical curves separating amplitude death regions from phase locking regions for different values of $\tau_1, \tau_2, \overline{\omega}$ and $\beta$.

In order to visualize these level curves, we plot $(\Delta, K)$ from equations (12) and (13) in terms of the variable $\beta$ for some specific values of the other parameters as shown in Fig. (2). Region (I) in the plot designates the region of solutions which evolve away from the unstable equilibrium at the origin towards phase locking, leading to stable limit cycles for $p_1(t)$ and $p_2(t)$. Their time series and phase space patterns are identical to those in fig. (1). Our analysis explains the associated flows of solutions between the unstable fixed point at $(0,0)$ and the equilibrium fixed point $(p_1^*, p_2^*)$ studied in [9,10].

On the other hand, region (II) represents solutions the evolve away from the unstable origin towards a stable fixed point that causes amplitude death solutions, with fixed limiting coherent values for $p_i(t)$ for the appropriate parameters. Fig. (2), on the right shows an amplitude death solution for $\tau_1 = \tau_2 = 2, \overline{\omega} = 5, \Delta = 4$ and $K = 10$. The prices rise from their initial conditions towards the equilibrium price, oscillate around it then fade down toward a fixed value afterwards corresponding to a stable equilibrium.

Even though our analysis is based on the linearized systems, the Hartman-Grobman theorem [13,14] guarantees that these results apply similarly on the non-linearized model in eq. (2). The theorem asserts that the dynamics of a system in a region around a hyperbolic equilibrium point exhibit the same qualitative behavior as its linearization near that equilibrium.

## 5. Higher Dimensions (N=4)

As these coupled equations represent the interaction between nonlinear oscillators, it is important to retain their phase and amplitude responses. Substituting $p_i = r_i e^{i\theta_i}$ and expanding the first term up to first order, the former equations could be expressed in polar form, the amplitude part is expressed by:

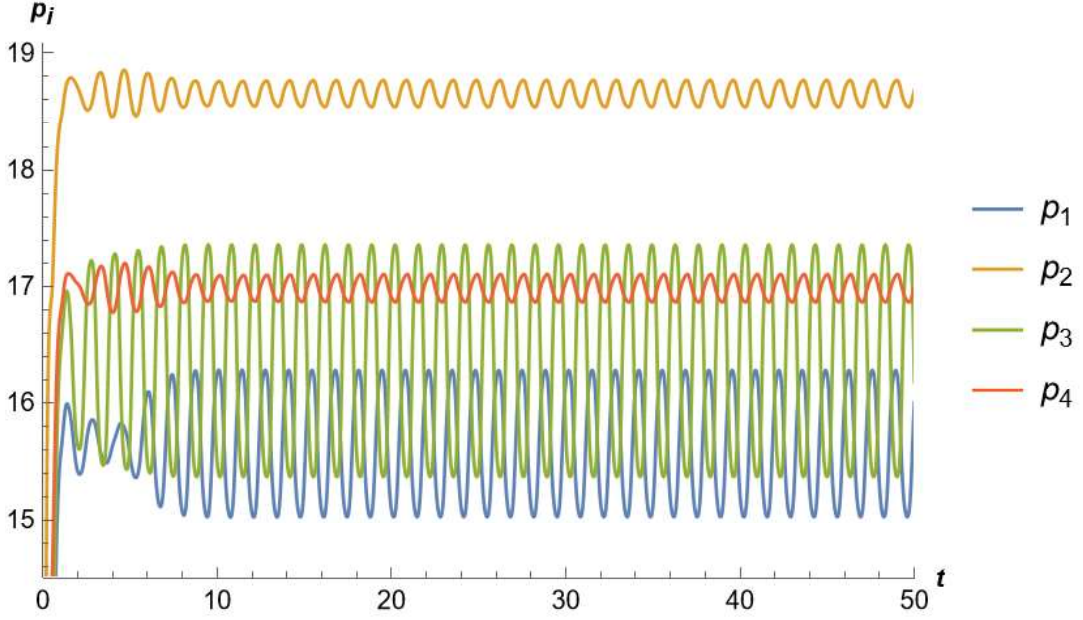

*Figure 3: A sample plot with limit cycles for an N = 4 coupled asset markets, with $m_1 = 0.5$, $m_2 = 0.65$, $m_3 = 0.4$, $m_4 = 0.65$, $\tau_i = 0.5\ \forall i$, $v_1 = 15$, $v_2 = 19.2$, $v_3 = 16$, $v_4 = 17$ and $K_{ij} = 1\ \forall i,j$.*

$$\dot{r}_i = \left[(1-2m_i)r_i\cos\theta_i - (1-m_i)r_i(t-\tau_i)\cos\theta_i(t-\tau_i) + m_i v_i - \sum_{j=1,j\neq i}^{N} K_{ij}\right]r_i$$

$$+ \sum_{j=1,j\neq i}^{N} K_{ij}\, r_j(t-\tau_j)\cos\bigl(\theta_j(t-\tau_j)-\theta_i\bigr) \tag{14}$$

while the phase part is given by:

$$\dot{\theta}_i = (1-2m_i)r_i\sin\theta_i + (1-m_i)r_i(t-\tau_i)\sin\theta_i(t-\tau_i)$$

$$+ \sum_{j=1,j\neq i}^{N} K_{ij}\frac{r_j(t-\tau_j)}{r_i}\sin\bigl(\theta_j(t-\tau_j)-\theta_i\bigr) \tag{15}$$

For *N*-asset systems, we can numerically verify that limit cycles can occur for systems with $N > 2$. Fig. (3) shows limit cycles for $N = 4$ with appropriate parameters and fig. (4) shows their corresponding phase diagrams $(\dot{p}_i, p_i)$ for $i = 1, \ldots, 4$ with clear synchronization of phase locking patterns. We also observe similar flow dynamics to the two-market system from prices with higher or lower initial values into supercritical cyclic oscillations around an unstable equilibrium point, when it exists.

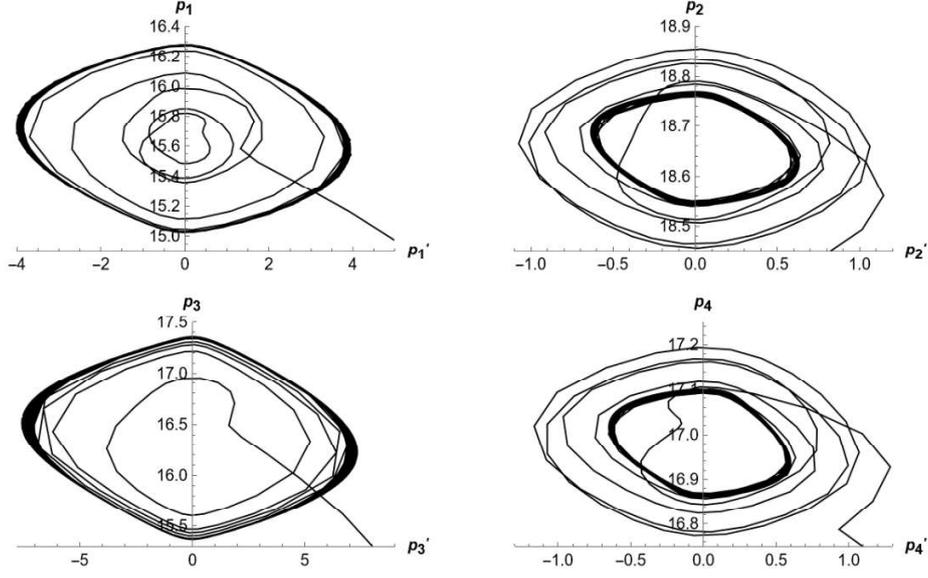

*Figure 4: Phase diagrams corresponding to the $N = 4$ coupled asset markets depicted in figure (3) with identical parameters.*

## 6. Generic N-asset model

In this section we introduce a generalized N-asset system given by:

$$\dot{p}_i = (1 - m_i) \tanh[p_i(t) - p_i(t - \tau)] p_i(t) - m_i(p_i(t) - v_i)p_i(t)$$
$$+ \frac{K'}{N} \sum_{j=1, j \neq i}^{N} \left(p_j(t - \tau) - p_i(t)\right) \quad (16)$$

where we assume that all time delays and market couplings are equal, such that $\tau_{ij} = \tau$ and $K_{ij} = K' = 2K$. We also consider the coupling strength to be inversely proportional to the market size $N$. Following the same analysis conducted in section I, we start by introducing the linearized matrix $S$ of the generic system given in (16) at $\dot{p}_i = 0$. Explicitly we find that

$$S = \begin{pmatrix} a_1 & f & \cdots & f \\ f & a_2 & \cdots & f \\ \vdots & \vdots & \ddots & \vdots \\ f & f & \cdots & a_n \end{pmatrix} \quad (17)$$

where $a_n = \omega_n - K'd$ and $f = \frac{K}{N} e^{-\lambda \tau}$ with $\omega_n = m_n v_n$ and $d = 1 - 1/N$.

More conveniently, we define a modified linearized matrix $M = S + Kd\, I$ with $I$ being the identity matrix. In this case, if $\mu$ is an eigenvalue of $M$, then it would be related to $\lambda$ by $\mu = \lambda + K'd$. The modified matrix $M$ is explicitly expressed as

$$M_{mn} = \begin{cases} \omega_n & m = n \\ f & m \neq n \end{cases} \qquad (18)$$

Its eigenvalues are determined by solving:

$$\det \begin{bmatrix} \omega_1 - \mu & f & \cdots & f \\ f & \omega_2 - \mu & \cdots & f \\ \vdots & \vdots & \ddots & \vdots \\ f & f & \cdots & \omega_n - \mu \end{bmatrix} = 0 \qquad (19)$$

Following the methods used in [15], the eigenvalue equation can be separated into a product of two factors that consecutively represent the continuous and the discrete spectra of the system. It is given by

$$\left[\prod_{k=1}^{N}(\omega_k - \mu - f)\right]\left[1 + f\sum_{j=1}^{N}\frac{1}{\omega_j - \mu - f}\right] = 0 \qquad (20)$$

This equation cannot be explicitly solved for any generic system, however, the case of $N$ oscialltors of identical frequencies provides analytical insights about the regions of amplitude death of the system.

*N-asset markets with identical frequencies*

We consider the special case of N-asset market with identical frequencies such that $\omega_i = m_i v_i = \omega_0\ \forall i$. The first factor in eq. (20) leads to the continuous spectrum solution, with the transcendental relation:

$$\lambda = \omega_0 - K'd - \frac{K'}{N}e^{-\lambda\tau} \qquad (21)$$

with $N - 1$ degeneracy in this case. While the second factor results in the discrete spectrum solution represented by the transcendental equation

$$\lambda = \omega_0 - K'd - K'd\, e^{-\lambda\tau} \tag{22}$$

This is a general solution for the eigenvalues of a system of an N-asset market with identical frequencies, coupling strengths and time delays. The bifurcations line, amplitude death regions and limit cycle regions could be obtained by similar analysis and numerical plots to those obtained in section four, for appropriate parameterization.

## 7. Conclusions

In this paper, we examined the interaction of a large number N of asset markets connected via linear time delay coupling, represented by a set of coupled delay differential equations. Our analysis demonstrated the existence of bifurcation, phase locking and synchronization regions in a two-asset market model. We determined the limit cycle and amplitude death regions by plotting the critical curves of the system's eigenvalues.

In higher dimensions, we pursued numerical investigations to confirm the existence of limit cycles in an N=4 system, serving as a prototype for higher-order asset markets. For a generic N asset market model, we established the existence of a general solution that specifies amplitude death and limit cycle regions, particularly in cases with identical oscillation frequencies.